\documentclass[amsmath,amssymb,aps,pra,reprint,groupedaddress]{revtex4-1}
\usepackage{graphicx}
\usepackage{dcolumn}
\usepackage{bm}

\begin{document}

%Title of paper
\title{Observation of Pendell\"{o}sung Fringes by Using Pulsed Neutrons}

\author{Shigeyasu Itoh}
\email[]{itoh@phi.phys.nagoya-u.ac.jp}
%\homepage[]{Your web page}
%\thanks{}
\author{Masaya Nakaji}
\author{Yuya Uchida}
\author{Masaaki Kitaguchi}
\author{Hirohiko M. Shimizu}
\affiliation{Department of Physics, Nagoya University\\
Furo-cho, Chikusa-ku, Nagoya, 464-8601, Japan}

\date{\today}

\begin{abstract}
Pendell\"{o}sung interference fringes of a single silicon crystal were observed by using pulsed cold neutrons. 
The nuclear scattering length of silicon was obtained as ($4.125\pm0.003(\rm{stat.})\pm0.028(\rm{syst.})$) fm using the observed Pendell\"{o}sung fringes. 
This indicates the applicability of pulsed neutron beam to observe the Pendell\"{o}sung fringes by using the time-of-flight analysis.
\end{abstract}

\maketitle

\section{Introduction}

Pendell\"{o}sung interference fringes in the X-ray diffraction of a crystal, resulting from dynamical diffraction in a perfect single crystal, were predicted by C. G. Darwin\cite{Darwin} and P. P. Ewald\cite{Ewald}. 
They were observed for the first time in 1959 by N. Kato and A. R. Lang\cite{Kato}.
Pendell\"{o}sung fringes of neutrons were also observed in 1968 by C. G. Shull\cite{Shull1}.

\begin{figure*}[t]
\centering
\begin{minipage}{1.8\columnwidth}
\includegraphics[clip, width=14cm, keepaspectratio]{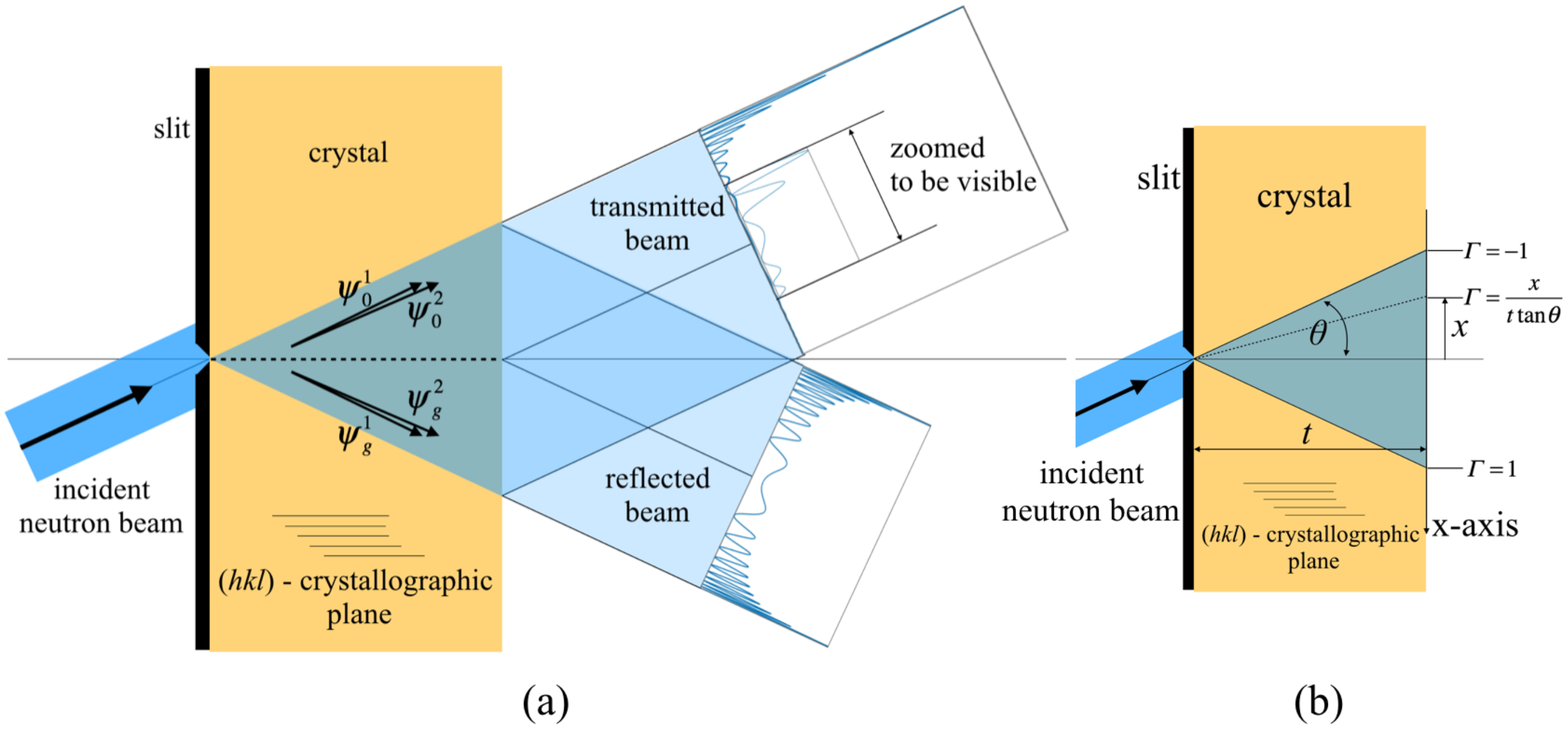}
\caption{(a) Neutron wave functions with symmetrical Laue geometry in a crystal can be written as a superposition of the four components by the dynamical diffraction theory. 
The incident neutrons propagate through the crystal spreading over a triangle zone (Borrmann fan) with a particular oscillating pattern due to the interference between the two Bloch waves in the transmitted and reflected beams, respectively. 
The curves on the right side represent the distributions of the emitted neutrons from the end surface of the crystal for the transmitted and reflected neutrons, respectively. 
In the transmitted direction, the intensity of the beam in the edge of the Borrmann fan is very strong, thus the central part is magnified to be visible. 
(b) Definition of the normalized position $\mathit\Gamma$ on the end surface of the crystal.}
\label{fig:fig1}
\end{minipage}
\end{figure*}

In the case of symmetrical Laue geometry (Fig.\ref{fig:fig1}(a)), when a neutron beam is injected into a perfect and thick crystal under the Bragg condition, the wave function in the crystal can be written as a superposition of the four components by the dynamical diffraction theory\cite{Sears, Lauch} as
%Equatio(1)
\begin{equation}
\psi=\psi_0^1+\psi_0^2+\psi_g^1+\psi_g^2 ,
\label{equation:1}
\end{equation}
where subscripts 0 and $g$ represent the transmitted and the reflected wave, respectively, and subscripts 1 and 2 represent two Bloch wave functions. 
The wavenumbers of these Bloch waves are slightly different corresponding to the periodical nuclear potential in the crystal. 
In general, the four components are coherent and can interfere with each other. The radiation density is given by
%Equation(2)
\begin{equation}
{|\psi|}^2\approx{|\psi_0^1+\psi_0^2|}^2+{|\psi_g^1+\psi_g^2|}^2 ,
\label{equation:2}
\end{equation}
as the intensity of $(\psi_0^1+\psi_0^2)^*(\psi_g^1+\psi_g^2)+(\psi_0^1+\psi_0^2)(\psi_g^1+\psi_g^2)^*$ oscillates rapidly to be smeared on propagation in the crystal.
The transmitted wave ${|\psi_0^1+\psi_0^2|}^2$ and the reflected wave ${|\psi_g^1+\psi_g^2|}^2$ generate a periodical structure of the radiation density in the
crystal. 
This structure is known as Pendell\"{o}sung interference fringes. 
Figure \ref{fig:fig1}(a) shows also the intensity distribution of neutrons on the end surface of the crystal, when neutrons is injected into the crystal through the narrow slit with the Bragg angle $\theta$ with respect to the crystallographic plane ($hkl$). 
The normalized distance $\mathit{\Gamma}\equiv x/(t\tan\theta)$ at the position $x$ from the diffraction center along the end surface would be introduced, where $t$ is the thickness of the crystal (see Fig.\ref{fig:fig1}(b)). 
%Figure 1
The intensity of the transmitted neurons at the position $\mathit{\Gamma}$ can be written as
%Equation(3)
%\begin{align}
\begin{widetext}
\begin{equation}
I_0(\mathit{\Gamma})d\mathit{\Gamma}={u_0}^2\mathit{\Delta}_0\tan\theta \frac{1-\mathit{\Gamma}}{\left(1+\mathit{\Gamma}\right)\sqrt{1-\mathit{\Gamma}^2}}\\\left[1+\cos\left(\frac{\pi}{2}+\frac{4tF_{hkl}d_{hkl}}{V_{\rm{c}}}\sqrt{1-\mathit{\Gamma}^2}\tan\theta\right)\right]d\mathit{\Gamma} ,
\label{equation:3}
%\end{align}
\end{equation}
\end{widetext}
where ${u_0}^2$ is the neutron density on the incident surface, $F_{hkl}$ and $d_{hkl}$ are the crystal structure factor and the spacing of the ($hkl$) crystallographic plane of the crystal, respectively, and $V_{\rm{c}}$ is the unit cell volume of the crystal. $\mathit{\Delta}_0$ is the Pendell\"{o}sung length, which is given by
%Equation(4)
\begin{equation}
\mathit{\Delta}_0=\frac{V_{\rm{c}}\pi\cos\theta}{\lambda F_{hkl}} ,
\label{equation:4}
\end{equation}
where $\lambda$ is the wavelength of the neutron. 
The neutron intensity through the exit slit for the transmitted beam, $P_0$ , can be obtained by integration of Eq.\ref{equation:3} as
%Equation(5)
\begin{equation}
P_0=\int_{-\mathit{\Gamma}_{\rm{s}}}^{\mathit{\Gamma}_{\rm{s}}}I_0(\mathit{\Gamma})d\mathit{\Gamma} .
\label{equation:5}
\end{equation}
In the case of the symmetric and narrow slit with the width of $2\mathit{\Gamma}_{\rm{s}}$ and $\it{\Gamma}\ll 1$ Eq.\ref{equation:5} can be expressed as
%Figure 2
\begin{figure*}[t]
\centering
\begin{minipage}{1.8\columnwidth}
\centering
\includegraphics[clip, width=12cm, keepaspectratio]{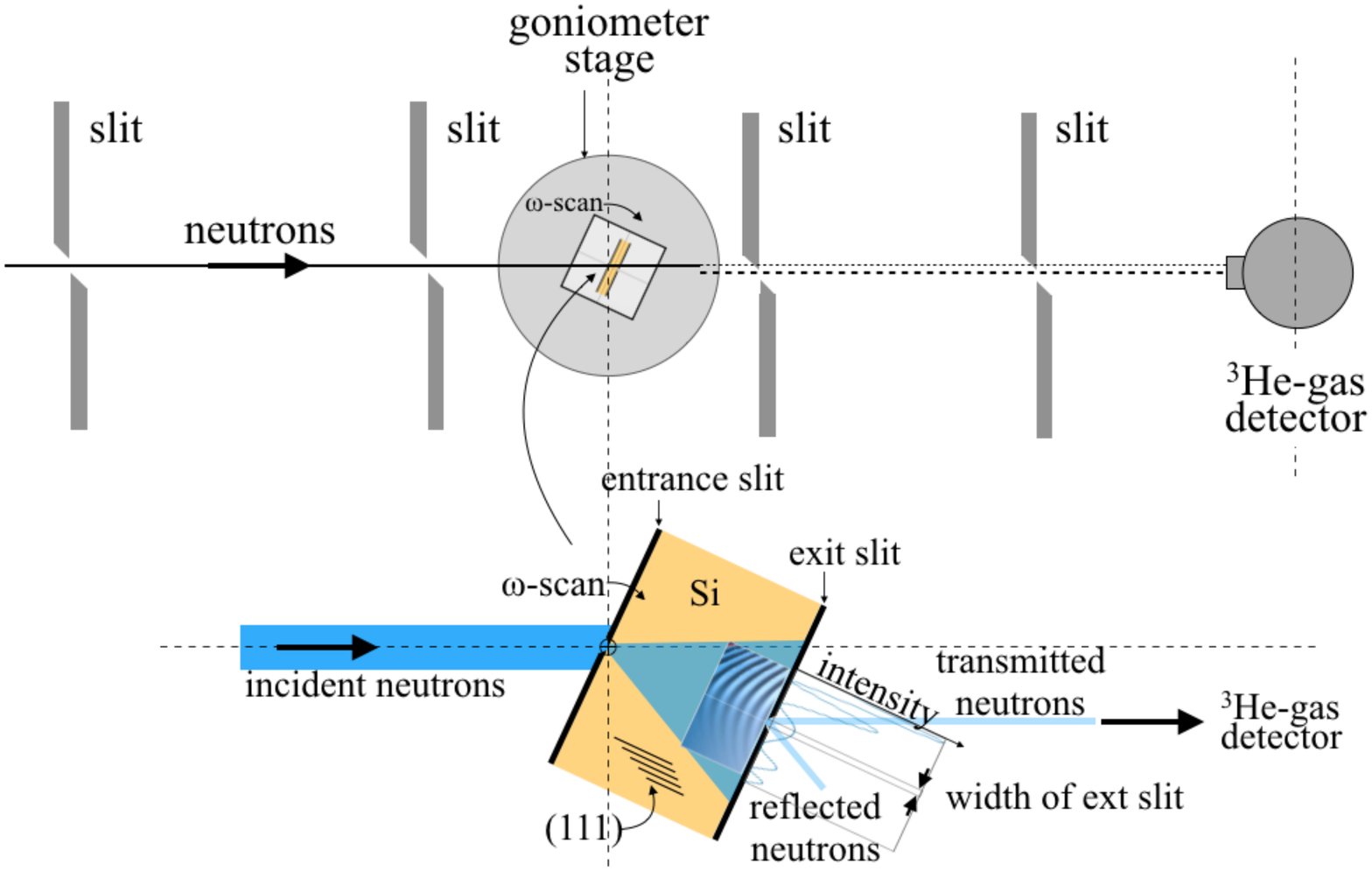}
\caption{Experimental setup to demonstrate Pendell\"{o}sung fringes at the MLF BL17 in J-PARC (top view). 
The goniometer stage was scanned clockwise in the range of $25.0^\circ - 26.5^\circ$. 
The real Bragg angle was obtained from the value of the TOF-position of the diffraction peak. 
The transmitted beam from the exit slit on the end surface of the crystal was shifted 1 mm downward in this figure and detected by the 3He-gas detector. 
The beam divergence was confined by the upstream two slits and entrance slit. 
Direct beam was cut off by the exit slit and the downstream two slits.}
\label{fig:fig2}
\end{minipage}
\end{figure*}
%
%Equation(6)
\begin{equation}
P_0\simeq 2{u_0}^2\mathit{\Delta}_0\mathit{\Gamma}_{\rm{s}}\tan\theta \left[1+\cos\left(\frac{\pi}{2}+\frac{4tF_{hkl}d_{hkl}}{V_{\rm{c}}}\tan\theta\right)\right] .
\label{equation:6}
\end{equation}
This periodic structure can be observed by varying $\theta$\cite{Shull2}. 
This can also be measured by varying the thickness of the crystal\cite{Somenkov, Kikuta}.
%Figure 3
\begin{figure}[t]
\centering
\begin{minipage}{0.9\columnwidth}
\centering
\centering\includegraphics[clip, width=6cm, keepaspectratio]{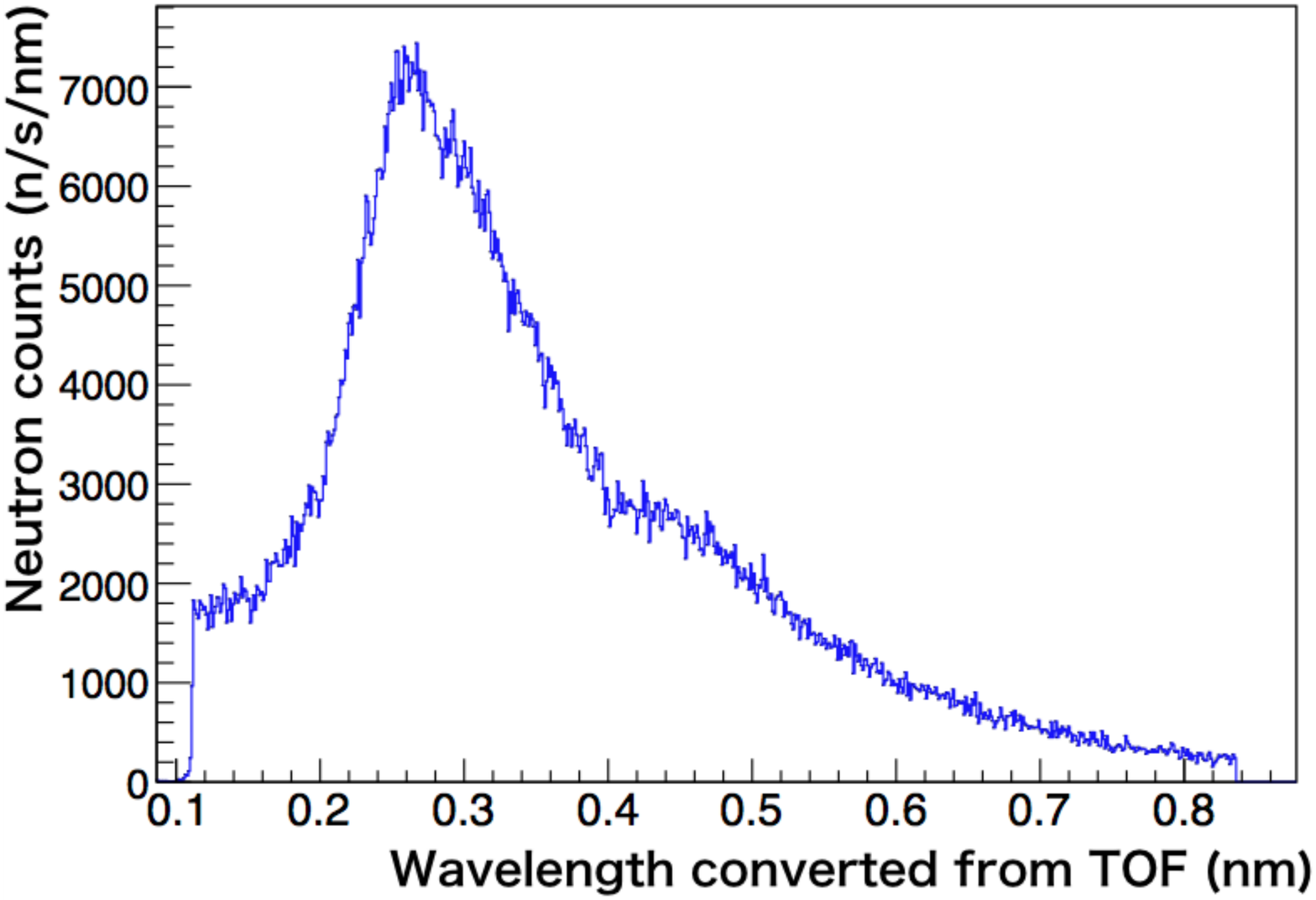}
\caption{Wavelength spectrum of the incident neutrons.}
\label{fig:fig3}
\end{minipage}
\end{figure}
The measurements of Pendell\"{o}sung fringes enable the accurate determination of the crystal structure factor, and then it can be used to obtain the value of the coherent scattering length of the atom. 
Using this method, Shull obtained a nuclear scattering length of $b_{\rm{nuclear}} = (4.1491 \pm 0.0010)$ fm for silicon\cite{Shull2}. 
These previous experiments were performed by using neutrons from reactors, therefore, monochromatic neutrons with the narrow energy-band was needed in order to satisfy the Bragg condition for a particular crystallographic plane.
In this paper, we report the demonstration of the applicability of pulsed neutron beam to observe the Pendell\"{o}sung fringes on several crystallographic planes at the same time by using the time-of-flight analysis.

\section{Experiment}

We used a single silicon crystal with a width of 50 mm, height of 50 mm, and thickness of 2.8 mm. 
The crystallographic plane (111) was used for demonstration. 
The surfaces of the crystal were mechanically polished and finally finished by chemical wet etching in order to remove the mosaic surface layer. 
The X-ray rocking curve exhibited a sharp peak with the width of less than 3 arcsec.
The experiment using pulsed neutrons was performed at the beam line BL17 in the Materials and Life Science Experimental Facility (MLF) in J-PARC (Japan Proton Accelerator Research Complex). 
Figure \ref{fig:fig2} shows the experimental setup. 
The crystal was sandwiched between the entrance cadmium slit and the exit cadmium slit which are 1.0 mm thick and have a 45$^\circ$tapering. 
The width of both the entrance and exit apertures was 0.2 mm. The neutron beam divergence was 0.039$^\circ$. 
The transmitted neutrons were measured by a $^3$He-gas detector with a gas pressure of 7 atm. 
The incident angle to the crystal was scanned by a goniometer from $\omega = 25.0^\circ$ to $\omega = 26.5^\circ$ with steps of $0.1^\circ$, where $\omega$ was the set angle of the goniometer stage. 
The measuring time of each step was two hours. 
Figure \ref{fig:fig3} shows the wavelength spectrum of the incident neutron beam.

\section{Results}

\begin{figure}[t]
\centering
\begin{minipage}{0.9\columnwidth}
\centering
\includegraphics[clip, width=8cm, keepaspectratio]{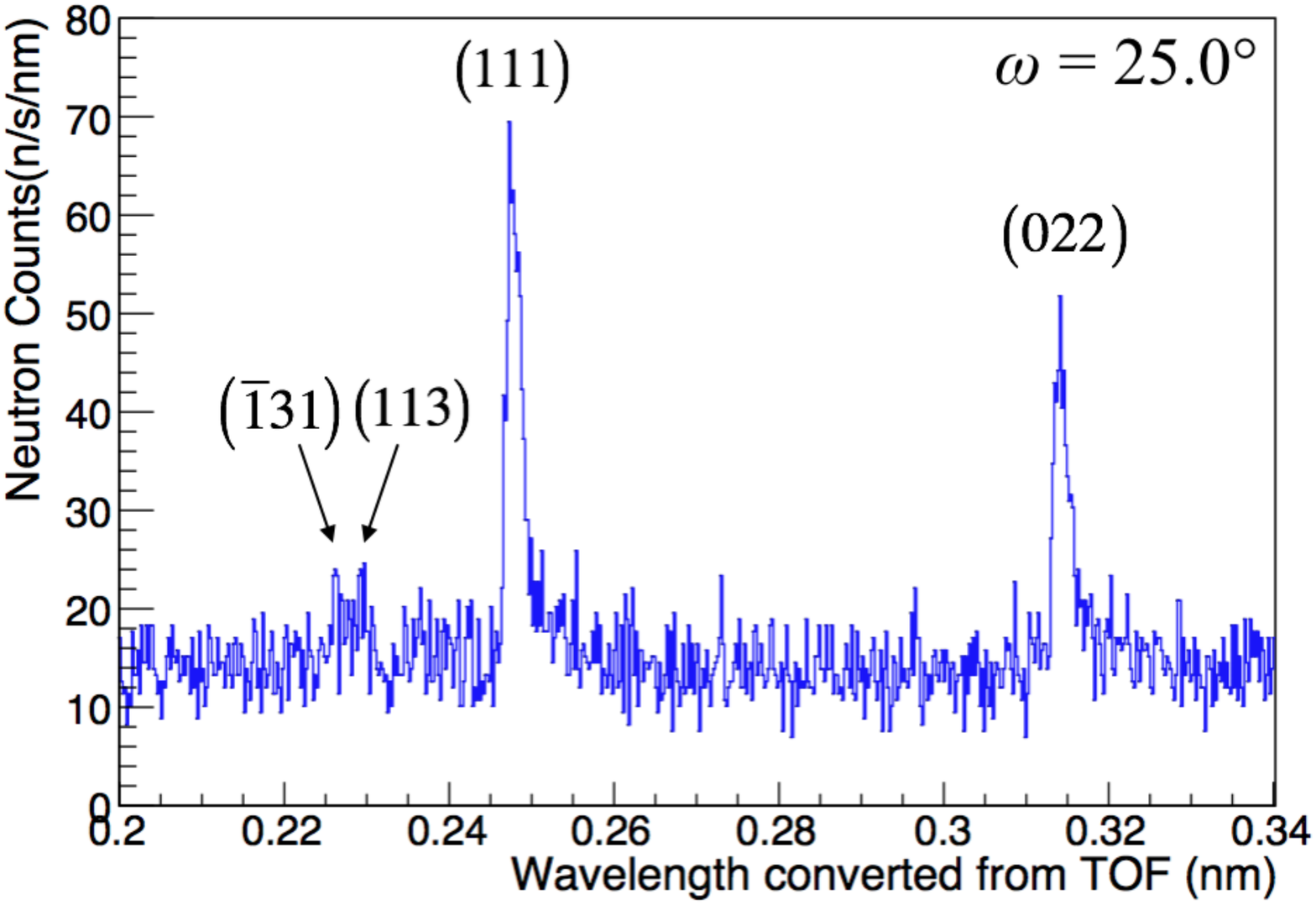}
\caption{Wavelength spectrum of the diffracted neutrons for the symmetrical Laue geometry of the (111) plane of a single silicon crystal for the set angle $\omega = 25.0^\circ$.}
\label{fig:fig4}
\end{minipage}
\end{figure}

Figure \ref{fig:fig4} shows the wavelength spectrum of the diffracted neutrons for $\omega = 25.0^\circ$. 
%Figure 4
%
%Figure 5
\begin{figure}[t]
\centering
\begin{minipage}{0.9\columnwidth}
\centering
\includegraphics[clip, width=6cm, keepaspectratio]{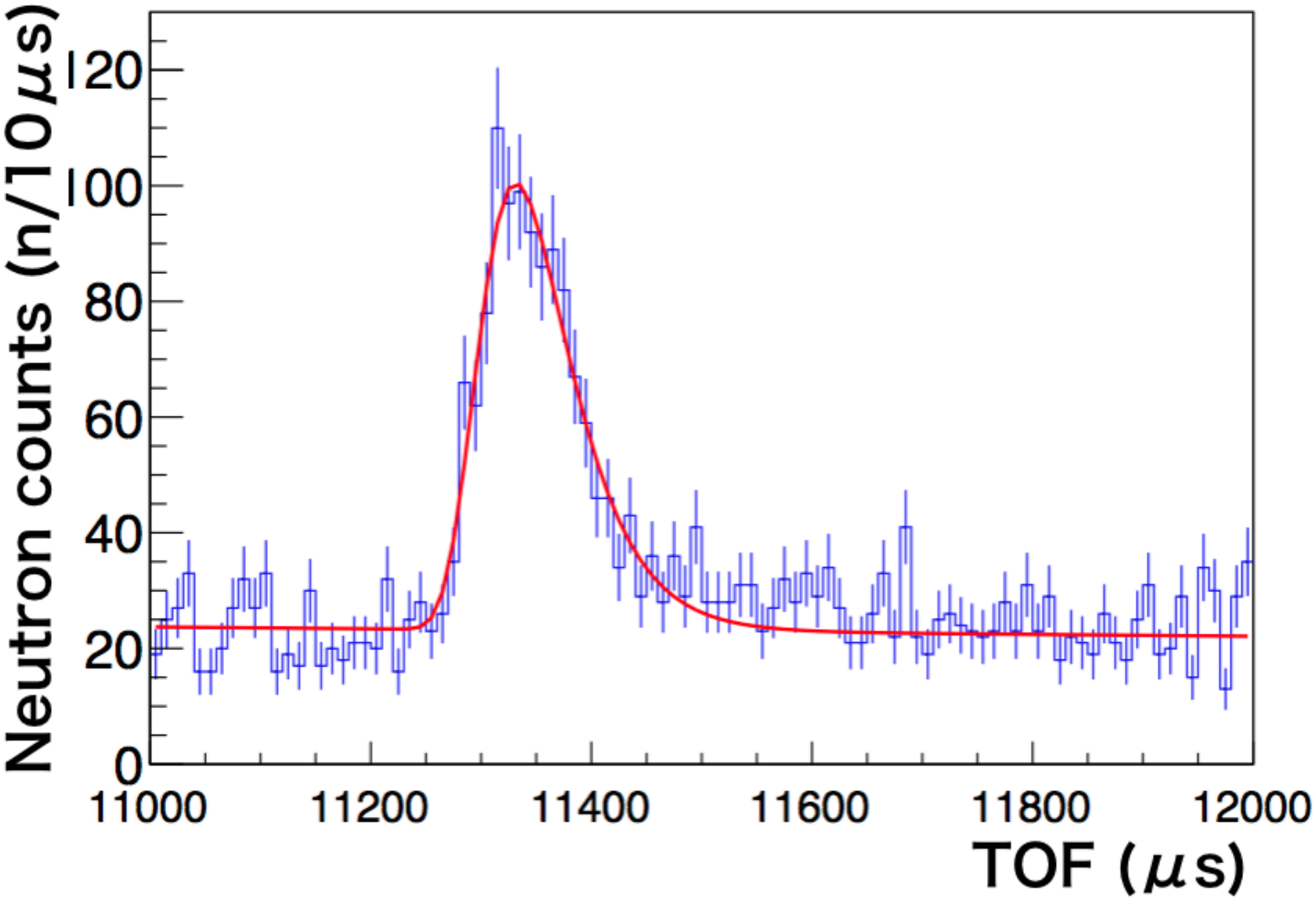}
\caption{Diffraction peak of the (111) plane for $\omega=25.0^\circ$ and the result of a fitting by the extreme function (Eq.\ref{equation:7}). The profile of the peak is fitted well.}
\label{fig:fig5}
\end{minipage}
\end{figure}
%
%Figure 6
\begin{figure}[t]
\begin{minipage}{0.9\columnwidth}
\centering
\includegraphics[clip, width=7cm, keepaspectratio]{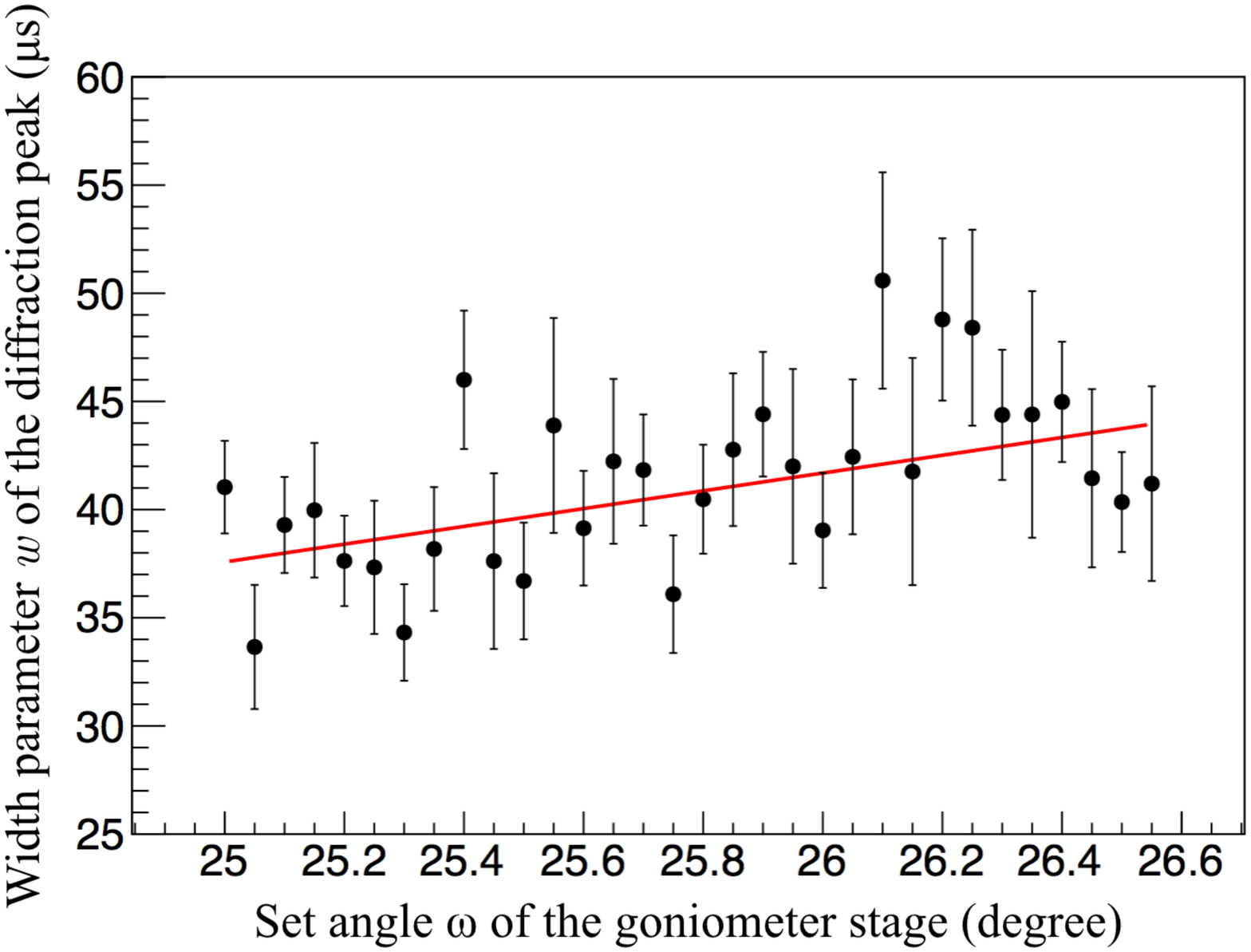}
\caption{Dependence of the width parameter $w$ on the goniometer angle $\omega$. The width of the diffraction peak tends to be broader as the wavelength is longer due to the characteristics of the cold}
\label{fig:fig6}
\end{minipage}
\end{figure}
Four diffraction peaks were observed for the (111) , (022) , (113) , (131) planes. 
The diffraction peak of the (111) plane, which was in a symmetrical Laue geometry condition, was analyzed for the demonstration of the Pendell\"{o}sung fringes. 
In order to determine the peak position and width, the peak in time-of-flight spectrum of diffracted neutrons was fitted by a combination of the linear function for the background and the exponential function for the diffraction peak as:
%Equation(7)
\begin{equation}
y=A+Bx+C\exp\left[-\exp\left(-\frac{x-x_{\rm{c}}}{w}\right)-\frac{x-x_{\rm{c}}}{w}+1\right] ,
\label{equation:7}
\end{equation}
where $A, B, C$ are constants (Fig.\ref{fig:fig5}). 
The exponential part of this function is known as the extreme function, where $x_{\rm{c}}$ is the maximum point corresponding to the peak position and $w$ represents the width of the peak. 
The width of the peak in TOF spectrum depends on the wavelength of the neutron due to the characteristics of the cold neutron source. 
The width parameter $w$ was corrected by using the fitting of its dependency on the goniometre angle (see Fig.\ref{fig:fig6}). 
The intensity of the diffraction peak at each scanning angle was defined as neutron counts in the full width of $2 \%$ of the maximum after the subtraction of the background. 
The Bragg angle $\theta$ was determined by the $TOF$ of the peak position using the relation;
%Equation(8)
\begin{equation}
TOF=\frac{m_{\rm{n}}Ld_{hkl}\sin\theta}{\pi\hbar} ,
\label{equation:8}
\end{equation}
where $m_{\rm{n}}$ is the mass of the neutron and $L$=18 m is the distance from the cold neutron source to the detector. 
The intensity of the diffraction peak was corrected for the glancing width of the cadmium slits and the spreading of the Borrmann fan at each incident angle. 
The intensity was also corrected for the incident beam spectrum and the energy width for the diffraction. 
The Pendell\"{o}sung fringes were clearly observed with a contrast of $(16\pm2) \%$ as shown in Fig.\ref{fig:fig7}.  
%Figure 7
\begin{figure}[t]
\centering
\begin{minipage}{0.9\columnwidth}
\centering
\includegraphics[clip, width=8cm, keepaspectratio]{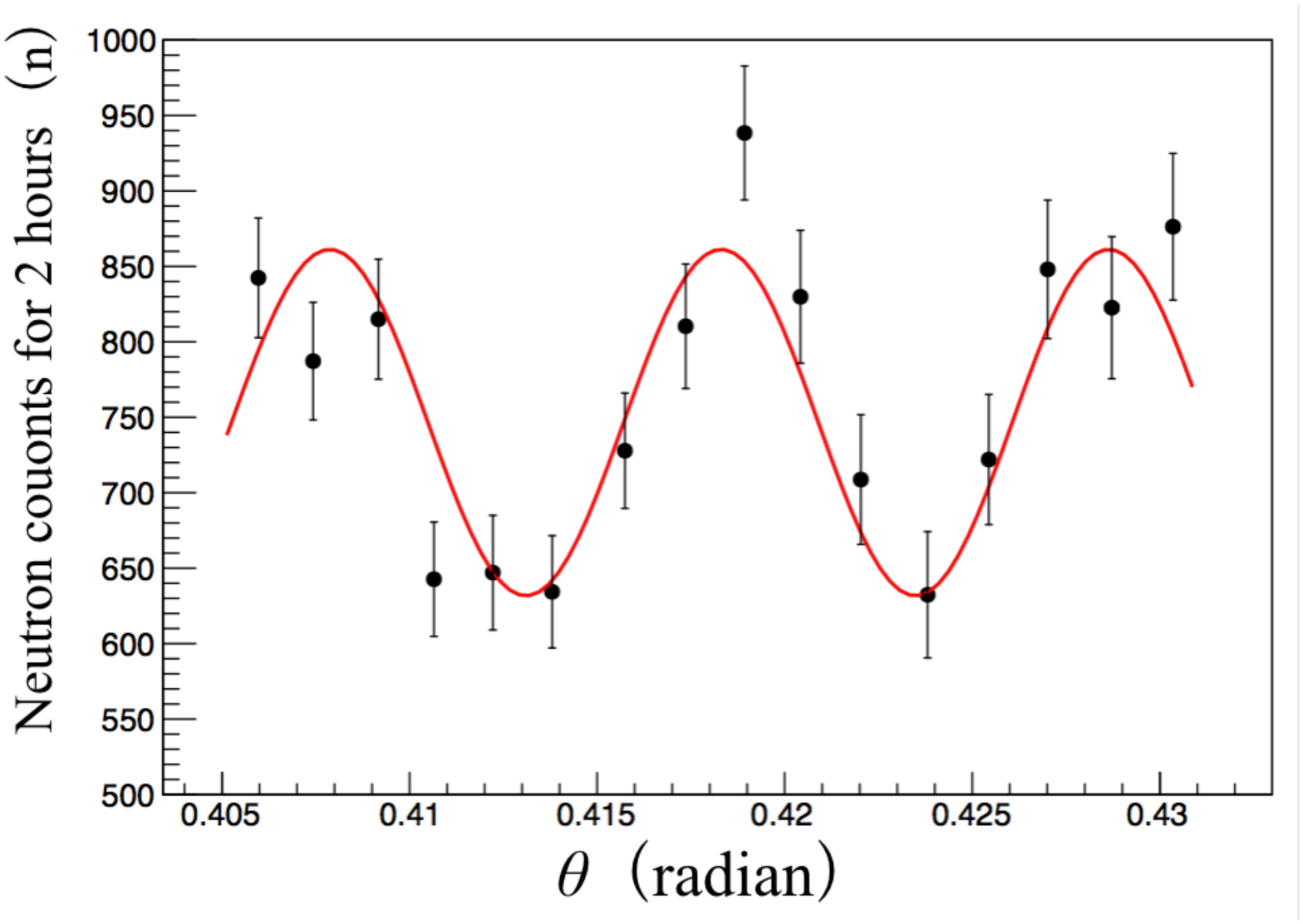}
\caption{Pendell\"{o}sung fringes of the (111) crystallographic plane of a single silicon crystal. The plots were fitted by Eq.\ref{equation:7}. The contrast was obtained at $\left(16 \pm 2\right) \%$.}
\label{fig:fig7}
\end{minipage}
\end{figure}
The coherent scattering length was obtained by the fitting with Eq.\ref{equation:6} of the Pendell\"{o}sung fringes. 
Silicon has a diamond structure, thus its $F_{hkl}$ for the (111) plane is given by
%Equation(9)
\begin{equation}
F_{111}=\sqrt{32}b_{\rm{c}}e^{-W} ,
\label{equation:9}
\end{equation}
where $b_{\rm{c}}$ is the coherent scattering length of a silicon and $e^{-W}$ is the Debye -Waller factor. 
We obtained
%Equation(10)
\begin{equation}
b_{\rm{c}}e^{-W}=\left(4.178\pm0.003\right)  (\rm{fm}) .
\label{equation:10}
\end{equation}
The Debye -Waller factor $e^{-W}$ of the (111) plane of the silicon was $(0.98842\pm0.00012)$\cite{Shull2}. 
The coherent scattering length $b_{\rm{c}}$ in Eq.\ref{equation:11} can be written as $b_{\rm{charge}} + b_{\rm{nuclear}}$, where $b_{\rm{charge}} $is the neutron-charge scattering length of $(+0.0043\pm0.0002)$ fm for the (111) plane of the silicon\cite{Shull2} and $b_{\rm{nuclear}}$ is the nuclear scattering length for the silicon. 
Consequently, we obtain
%Equation(11)
\begin{equation}
b_{\rm{nuclear}}=\left(4.125\pm0.003(\rm{stat.})\pm0.028(\rm{syst.})\right)  (\rm{fm}) .
\label{equation:11}
\end{equation}
The systematic uncertainties and the statistical uncertainty are summarized on Tab.\ref{tab:uncertainty}. 
This value was consistent with the value of $(4.1491\pm0.0010)$ fm from the National Institute of Science and Technology (NIST). 
The origin of the largest systematic uncertainty is the Bragg angle $\theta$, which was derived from the accuracy of the wavelength of the incident neutron.
%Table 1
\begin{table*}[t]
\centering
\begin{minipage}{1.8\columnwidth}
\centering
\caption{Systematic and statistical uncertainties\newline
* Calculated from the lattice constant $a_0 =0.5431020504(89)$ fm from NIST.\newline
** Referenced from \cite{Shull2}}
\small
\begin{tabular}{l|ccccc} \hline
Item & Symbol & Unit & Value & Uncertainty & Uncertainty of $b_{\rm{nuclear}}$ \\ \hline \hline
Systematic uncertainty & & & & & \\
Bragg angle & $\theta$ & radian & 0.4006 & 0.00241 & 0.02755 \\
Thickness of crystal & $t$ & mm & 2.795 & 0.003 & 0.0045 \\
Cell volume & $V_{\rm{c}}$ & $\rm{nm}^3$ & $^*0.1602$ & 0.0009 & 0.000023 \\
Spacing of (111)-plane & $d_{hkl}$ & nm & $^*0.3135601$ & 0.000006 & 0.000008 \\
Debye-Waller factor & $e^{-W}$ & - & $^{**}0.98842$ & 0.00012 & 0.00051 \\
Total & & & & & 0.028 \\ \hline
 Statistical uncertainty & & & & & 0.003 \\ \hline
\end{tabular}
\label{tab:uncertainty}
\end{minipage}
\end{table*}

\section{Conclusion}

Pendell\"{o}sung interference fringes using pulsed cold neutrons were observed for the first time. 
The observed fringes can be explained with the dynamical diffraction theory for the crystal. 
The coherent scattering length was accurately measured by using the period of the fringes. 
The time-of-flight analysis for a pulsed neutron beam enables the use of polychromatic neutrons for the observations of Pendell\"{o}sung fringes with respect to several crystallographic planes at the same time. 
We expect that the simultaneous measurements on the planes would introduce an improved accuracy of physics observables that can be determined in their ratio.

\section*{Acknowledgment}

The neutron scattering experiment was approved by the Neutron Scattering Program Advisory Committee of IMSS, KEK (Proposal No. 2014S03). 
The neutron experiment at the Materials and Life Science Experimental Facility of the J-PARC was performed under a user program (Proposal No. 2015AU1701, 2016A0227, 2016B0141, 2017A0063). 
This work was supported by MEXT KAKENHI Grant Number JP19GS0210 and JSPS KAKENHI Grant Number JP16K13804.

\end{document}